\documentstyle[12pt]{article}
\begin{document}
\setlength{\baselineskip}{0.42in}

\newcommand{\nc}{\newcommand}
\nc{\beq}{\begin{equation}}   \nc{\eeq}{\end{equation}}
\nc{\beqa}{\begin{eqnarray}}  \nc{\eeqa}{\end{eqnarray}}
\nc{\lsim}{\begin{array}{c}\,\sim\vspace{-21pt}\\< \end{array}}
\nc{\gsim}{\begin{array}{c}\sim\vspace{-21pt}\\> \end{array}}
\nc{\eg}{E_\gamma}  \nc{\cg}{\cos\theta_\gamma}

\begin{titlepage}
\begin{center}
\setlength{\baselineskip}{0.25in}
{\hbox to\hsize{April 1994 \hfill hep-ph/9405216}}
{\hbox to\hsize{JHU-TIPAC-940004 \hfill UM-TH-94-16}}

\vskip 1.0 cm
\setlength{\baselineskip}{0.45in}
{\Large \bf Bounding the Tau Neutrino Magnetic Moment}\\
{\Large \bf From Single Photon Searches at LEP}
\vskip 1.0 cm

\begin{tabular}{c}
{\bf Thomas M. Gould\footnotemark[1]}\\[.05in]
{\it Department of Physics and Astronomy}\\
{\it The Johns Hopkins University}\\
{\it Baltimore MD 21218 }\\[.15in]
\end{tabular} \\
\begin{tabular}{c}
{\bf I.Z. Rothstein\footnotemark[2]}\\[.05in]
{\it The Randall Laboratory of Physics}\\
{\it University of Michigan}\\
{\it Ann Arbor MI 48109 }\\[.15in]
\end{tabular}
\vskip 0.25cm
\footnotetext[1]{gould@fermi.pha.jhu.edu}
\footnotetext[2]{irothstein@umiphys.bitnet}

{\bf Abstract}\\[-0.05in]

\setlength{\baselineskip}{0.35in}
\begin{quote}

We show that single photon searches at LEP constrain
the tau neutrino magnetic moment to be less than
${\cal O}(10^{-6})~\mu_B$.
This bound is competitive with low energy ($s\simeq (30~GeV)^2$)
single photon searches.

\end{quote}
\end{center}
\end{titlepage}
\newpage
\setlength{\baselineskip}{0.35in}
The present bounds on the electromagnetic properties of the tau
neutrino are several orders of magnitude worse than the bounds on
the electron and muon neutrinos.
To date, the best bounds on tau neutrino electromagnetic properties
have been derived from single photon searches
at $e^+e^-$ colliders below the $Z$ resonance,
$\sqrt{s} \simeq 30~GeV$  (e.g.-- MAC, ASP, CELLO,
\mbox{Mark-J)}~\cite{grotch}.
In this short note,
we point out that comparable constraints on the tau neutrino
magnetic moment can be obtained from single photon searches at LEP.
These bounds are relevant for ruling out the possibility of
the tau neutrino having a mass in the $MeV$ range~\cite{BGR,dgt}.

Single photon searches at L3 have provided a measurement
of the number of neutrino species ($N_\nu = 3.24 \pm 0.46 \pm 0.22$),
as well as bounds on the couplings of heavy ($m_\nu* > 43~GeV$)
``excited'' neutrino to electroweak gauge particles ($E_\gamma >
10~GeV$)~\cite{L31},
and the compositeness of the $Z$ and the mass of light gravitinos
($E_\gamma > \frac{1}{2}E_{beam}$)~\cite{L32}.
Single photon searches at ALEPH have also provided a measurement
of the number of neutrino species
as well as bounds on the couplings of heavy ``excited'' neutrino
to electroweak gauge particles~\cite{ALEPH2}.
Rizzo has derived constraints on anomalous $Z\nu\bar{\nu}$ couplings
from the L3 measurement of the $Z$ width~\cite{RIZZO}.
In a comprehensive approach,
Escribano and Mass\'o  have recently derived constraints on
all electromagnetic fermion couplings from LEP-I data~\cite{EM}.
We show below that the LEP searches for  single energetic photons,
with no other particles detected,  also provide
a strong accelerator-based constraint on the tau neutrino magnetic moment.

Possible electromagnetic properties of a massive Dirac neutrino
are summarized in the current
\beq
J_\mu ~=~ e~\bar{\nu}(p^\prime)~\left(\, F_1(q^2)~\gamma_\mu ~+~
i~{F_2(q^2)\over 2\, m_\nu}~\sigma_{\mu\nu}~q^\nu \,\right) ~\nu(p)
\eeq
with $F_{1,2}(q^2\equiv (p^\prime - p)^2)$ dimensionless structure functions.
The magnetic moment $\kappa$ is
\beq
\kappa~\mu_B ~=~ {e\, F_2(0)\over 2\, m_\nu} ~,
\eeq
in units of the Bohr magneton $\mu_B$.
Single photon searches probe the form factor at zero momentum transfer,
obviating off-shell extrapolations.

At low center of mass energy $s\ll M_z^2$,
the dominant contribution to the process
$$e^+~e^-~\rightarrow~\nu~\bar{\nu}~\gamma$$
involves the exchange of a virtual photon~\cite{grotch}.
Dependence on the magnetic moment comes from a direct coupling
to the virtual photon,
and the observed photon is a result of initial state {\em Bremsstrahlung}.
At higher $s$, near the $Z$ pole $s \simeq M_z^2$,
the dominant contribution for $E_\gamma > 10~GeV$
involves the exchange of a $Z$.
Dependence on the magnetic moment now comes from the final state
radiation of the observed photon.
We emphasize here the importance of final state radiation
near the $Z$ pole,
which occurs preferentially at high $E_\gamma$ compared to conventional
{\em Bremsstrahlung}.
It has been previously believed that physics dominated by the
$Z$ pole is insensitive to electromagnetic properties~\cite{grotch}.

The magnetic moment coupling gives a contribution to the
differential cross-section for the process
$e^+e^-\rightarrow \nu\bar{\nu}\gamma $ of the form
\beq
\label{cs}
{d\sigma\over \eg d\eg~d\cg} ~=~
{\alpha^2\kappa^2\over 96\pi}~\mu_B^2~C\left[\, x_w \,\right]~
F\left[\, s,\eg,\cg \,\right]~
\eeq
where $\eg,\cg$ are the energy and scattering angle of
the photon.
The kinematics are contained in the function
\beq
\label{f}
F\left[\, s,\eg,\cg \,\right]  ~\equiv~
{s \, - \, 2\,\sqrt{s}\,\eg \, + \, {1\over 2}\,\eg^2\,\sin^2\theta_\gamma
\over \left(s - M_z^2\right)^2 \, + \, M_z^2\Gamma_z^2}
{}~.
\eeq
The coefficient $C$ is
\beq
\label{c}
{\cal C}\left[\, x_w \,\right]  ~\equiv~
{8\, x_w^2 \, -\, 4\, x_w \, +\, 1 \over x_w^2\, \left(1 - x_w\right)^2}~,
\eeq
using Standard Model $Ze\bar{e}$ and $Z\nu\bar{\nu}$ couplings,
and $x_w\equiv \sin^2\theta_w$.
In using (\ref{cs}), we neglect initial state radiation,
and $W$ and photon exchange graphs, which amount to $1\%$ corrections
in the relevant kinematic regime.

Now we constrain the magnetic moment by combining some LEP data sets,
with different integrated luminosities $L$, cuts on photon angle
and energy. There were no events allowed by the cuts in any of the
experiments.

\bigskip
\begin{center}
\renewcommand{\arraystretch}{1.2}
\begin{tabular}{||ll|c|c|c|c|c|} \hline
Expt. &  & $L \{pb^{-1}\}$ & $\theta_{min}$,$\theta_{max}$ &
$E_{min} \{GeV\}$ & Events & ref. \\ \hline
L3    & 1990 & $3.8 $  & $45$,$135$ & $10$   & $0$ & \cite{L31} \\ \cline{3-7}
      & 1991 & $11.2$  & $45$,$135$ & $22.8$ & $0$ & \cite{L32} \\ \hline
ALEPH & 1991 & $8.5 $  & $42$,$138$ & $17$   & $0$ & \cite{ALEPH2}\\ \hline
\end{tabular}
\end{center}
\smallskip
\centerline{Table 1}

Integrating (\ref{cs}) over the relevant range for each experiment,
we obtain estimates of the number of single photon events.
The combined number of events expected at
LEP due to a neutrino magnetic moment is
\beq
\label{number}
N ~\simeq~ 7.2 ~ 10^{10} ~\kappa^2 ~.
\eeq
from data at the $Z$ resonance {\it only}.
Using Poisson statistics,
we require (\ref{number}) be less than $2.2$,
giving  a bound on the magnetic moment at the 90\% confidence level.
This implies a bound
\beq
\label{thebound}
\kappa ~\lsim~ 5.5 ~10^{-6} \hspace{1.0cm} {\rm at~90\%~ C.L.}
\eeq

This result compares favorably with the bounds obtained from low-energy
single photon searches ($4.0~10^{-6}$ at 90\% C.L.)~\cite{grotch} and from
LEP-I data on $Z$ partial widths ($3.6~10^{-6}$ at 68\% C.L.)~\cite{EM}.
The derived  bound (\ref{thebound})  could be improved
by including data from the entire $Z$ resonance and may supercede the best
bound in ~\cite{EM}.
We estimate that the inclusion of the full integrated luminosity for each
experiment can at best improve the bound by $\sqrt{2}$, giving
\beq
\label{bestbound}
\kappa ~\lsim~ 3.9 ~10^{-6} \hspace{1.0cm} {\rm at~90\%~ C.L.}
\eeq
We expect a more careful analysis of the data will yield a bound between
(\ref{thebound}) and (\ref{bestbound}).

The bound applies to Dirac as well as Majorana transition moments.
However, transition moments involving the electron and muon flavors
are more strongly  bounded  by other accelerator experiments
($\kappa_e = 1.1~10^{-9}$,  $\kappa_\mu = 7.4~10^{-9}$)~\cite{emubounds}.
More stringent bounds on neutrino magnetic moments ${\cal O}(10^{-(10-12)})$
are available from astrophysical constraints~\cite{ASTRO},
but these bounds are apply only to neutrino masses which do not exceed
stellar temperatures ($T\simeq {\cal O}(MeV)$).
Furthermore, astrophysical constraints are ``model dependent''
in that they assume no new physics beyond the standard model~\cite{bmr}.
Elastic scattering experiments also give more stringent bounds, but
these bounds are mass dependent and become quite weak for
light neutrinos~\cite{BGR}.

The magnetic moment bound is relevant for ruling out
MeV tau neutrinos since MeV neutrinos must decay rapidly to avoid
nucleosynthesis constraints. In ref. \cite{BGR,dgt},
it was pointed out that there is an open window for radiative decay
into a sterile species.
This window is further closed by the constraints discussed above.

\centerline{\bf Acknowledgements}
\smallskip
TMG would like to thank P. Fisher for useful discussions.
The authors would like to acknowledge the hospitality of the ITP
at Santa Barbara where some of this work was done.

\nc{\ib}[3]{        {\em ibid. }{\bf #1} (19#2) #3}
\nc{\np}[3]{        {\em Nucl. Phys. }{\bf #1} (19#2) #3}
\nc{\pl}[3]{        {\em Phys. Lett. }{\bf #1} (19#2) #3}
\nc{\pr}[3]{        {\em Phys. Rev.  }{\bf #1} (19#2) #3}
\nc{\prep}[3]{      {\em Phys. Rep.  }{\bf #1} (19#2) #3}
\nc{\prl}[3]{       {\em Phys. Rev. Lett. }{\bf #1} (19#2) #3}


\newpage
\baselineskip=0.35in
\begin{thebibliography}{99}
\bibitem{grotch} H. Grotch and R. Robinett,
{\em Z. Phys.} {\bf C39} (1988) 553.
\bibitem{BGR}K.S. Babu, T.M. Gould and I.Z. Rothstein, \pl{B321}{94}{140}.
\bibitem{dgt}S. Dodelson, G. Gyuk and M. Turner, FERMILAB-Pub-93236-A
\bibitem{L31}B. Adeva et al., L3 collaboration, \pl{252B}{90}{527}.
\bibitem{L32}O. Adriani et al., L3 collaboration, \pl{297B}{92}{471};
\prep{236}{93}{1}.
\bibitem{ALEPH2}D. Decamp et al., ALEPH collaboration, \prep{216}{92}{253}.
\bibitem{RIZZO}T. Rizzo, \pl{237B}{90}{88}.
\bibitem{EM}R. Escribano and E. Mass\'o, Aut\`onoma de Barcelona preprint
UAB-FT-317, hep-ph/9403304.
\bibitem{emubounds}
D.A. Krakauer et al., LAMPF collaboration, \pl{252B}{90}{177},
\pr{44D}{91}{6}.
\bibitem{ASTRO}
G. Raffelt,\prl{64}{90}{2856};
G. Raffelt, D. Dearborn and J. Silk, {\em Ap. J.} {\bf 336} (1989) 61;
M. Fukugita and S. Yazaki, \pr{D36}{87}{3817};
S. Nussinov and Y. Rephaeli, \pr{D36}{87}{2278};
M.A.B. Beg, W. Marciano and M. Ruderman, \pr{D17}{78}{1395}.
\bibitem{bmr}K.S. Babu, R.N. Mohapatra and I.Z. Rothstein \pr{D45}{92}{R3312}
\end{thebibliography}
\end{document}